\documentclass[conference,onecolumn]{IEEEtran}\IEEEoverridecommandlockouts

\usepackage{cite}
\usepackage{amsmath,amssymb,amsfonts}
\usepackage{algorithmic}
\usepackage{graphicx}
\usepackage{textcomp}
\usepackage{xcolor}
\def\BibTeX{{\rm B\kern-.05em{\sc i\kern-.025em b}\kern-.08em
    T\kern-.1667em\lower.7ex\hbox{E}\kern-.125emX}}

\begin{document}

\title{An Ultra-wideband Battery-less Positioning System\\ for Space Applications\\
\thanks{This work has been funded by the European Space Agency, under contract number ESA AO 1-8471/15/NL/LvH}
}

\author{\IEEEauthorblockN{Davide Dardari, Nicol\`o Decarli, Davide Fabbri, Anna Guerra, Marco Fantuzzi, Diego Masotti, Alessandra Costanzo,\\ Aldo Romani} 
\IEEEauthorblockA{\textit{DEI-CNIT,}
\textit{Universit\`a degli Studi di Bologna}, 
Cesena, Italy -  {\it \{name.surname\}}@unibo.it}
\and
\IEEEauthorblockN{Maxime Drouguet, Thomas Feuillen, Christopher Raucy, Luc Vandendorpe, Christophe Craeye}
\IEEEauthorblockA{\textit{ICTEAM Institute,}
\textit{Universit\'e catholique de Louvain}, 
Louvain-la-Neuve, Belgium - 
maxime.drouguet@uclouvain.be}
}


\IEEEoverridecommandlockouts
\IEEEpubid{\begin{minipage}{\textwidth}\ \\[10pt]
\centering\normalsize{978-1-7281-0589-5/19/\$31.00 \copyright 2019 IEEE, \\ \tiny{Personal use of this material is permitted.  Permission from IEEE must be obtained for all other uses, in any current or future media, including reprinting/republishing this material for advertising or promotional purposes, creating new collective works, for resale or redistribution to servers or lists, or reuse of any copyrighted component of this work in other works.}}
\end{minipage}} 

\maketitle

\begin{abstract}
An ultra-wide bandwidth (UWB) remote-powered positioning system for potential use in tracking floating objects inside space stations is presented.
It makes use of battery-less tags that are powered-up and addressed through wireless power transfer in the UHF band and  
embed an energy efficient pulse generator in the 3-5 GHz UWB band.
The system has been mounted on the ESA Mars Rover prototype to demonstrate its functionality and performance. 
Experimental results show the feasibility of centimeter-level localization accuracy at distances larger than 10 meters, 
with the capability of determining the position of multiple tags using a 2W-ERP power source in the UHF RFID frequency band.
\end{abstract}


\section{Introduction}

In recent years we have assisted to an increasing interest in localizing objects and persons equipped with low-cost and battery-less tags in indoor environments to enable several applications in different fields  such as logistics, retail, security, etc.   
\cite{MieEtAll:11}.

UHF Gen.2 radio-frequency identification (RFID) standard technology is currently the most popular solution for item identification.  Unfortunately, it  has been designed having in mind identification but not positioning so that only rough position information can be obtained with commercial readers. 
Some approaches have been proposed to improve the localization accuracy \cite{NiZha:11}, but  they are typically not reliable in harsh propagation environments or require expensive hardware at reader side (e.g., large antenna arrays).

\begin{figure}[t]
\centerline{\includegraphics[width=0.6\columnwidth]{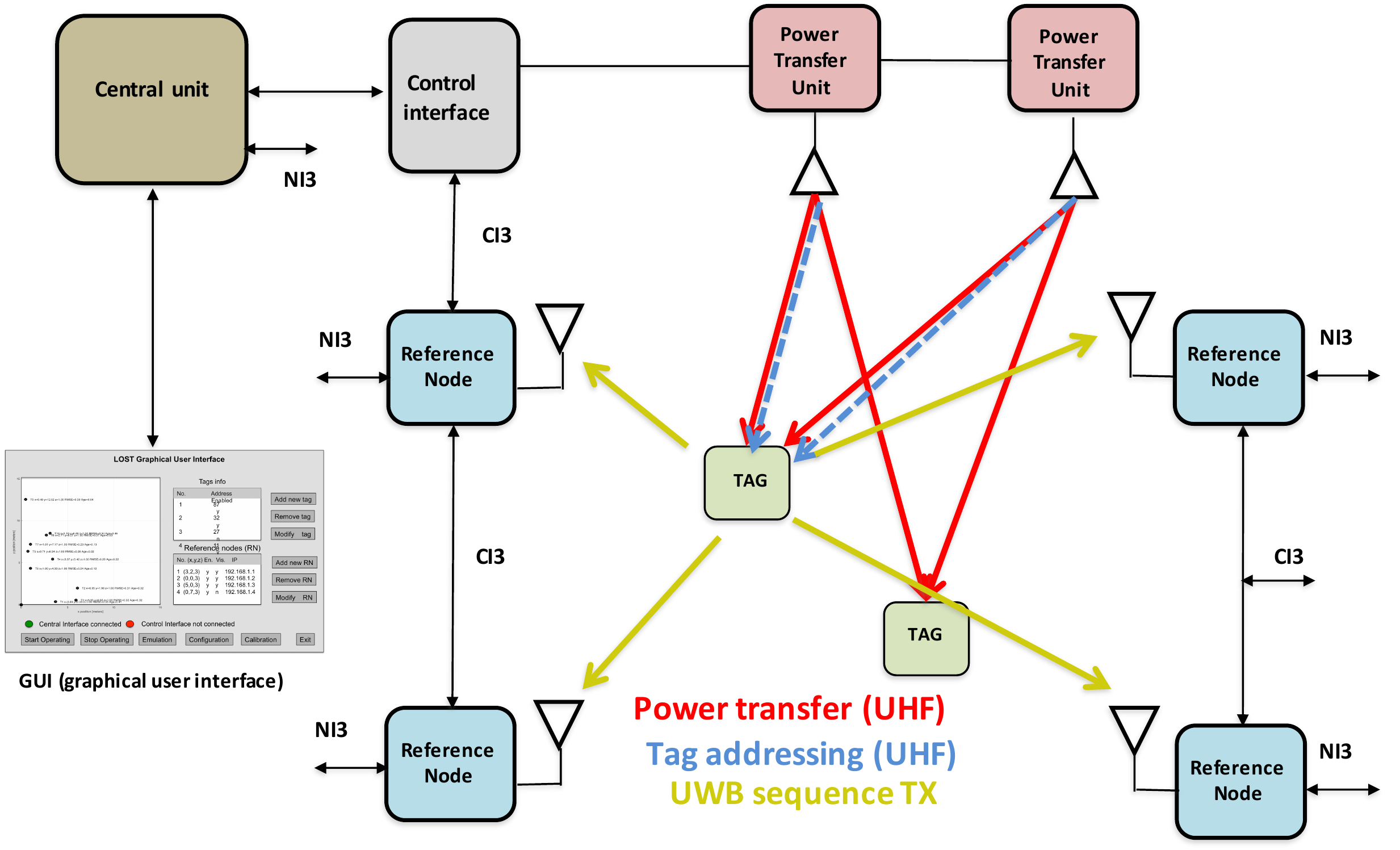}}
\caption{Architecture of the LOST system.}
\label{Fig:Scheme}
\end{figure}

At the same time, some new real-time locating systems (RTLS)  have emerged providing high-accuracy localization by adopting ultra-wide bandwidth (UWB) signals and exploiting their fine time resolution capabilities \cite{DarCloDju:J15}. Nevertheless, current UWB-based localization systems make use of active tags with a current consumption larger than $50\,$mA, which is not compatible with the exploitation of energy harvesting or wireless power transfer techniques, so that batteries or extremely low duty cycle operations are unavoidable \cite{RuiGra:17}. 

 \IEEEpubidadjcol

Recently, following the same backscattering philosophy as in standard Gen.2 RFID systems, some solutions have been proposed to realize battery-less tags working with UWB backscattered signals achieving interesting results in terms positioning accuracy (about 5-15 cm) \cite{DarDErRobSibWin:J10,DerKeiRud:13, 
DecGuiDar:J16, ArnMueWit:10,AleDecGueDar:C17, DErAl:C12, DarDecGueGui:C16,CosDarAleDecDelFabFanGueMasPizRom:J17}.
Despite its promising characteristics in terms of low-complexity and low-power consumption, 
the backscatter-based architecture suffers from strong link-budget (2-way link due to the reflected signal) 
which, in addition to the very conservative regulatory power emission constraints in the UWB band, limits 
its application only to very-short range scenarios (coverage range $<10$ meters) \cite{DecDar:J18}.

In this paper we illustrate a RTLS with battery-less tags capable of extending the range even beyond 10 meters by using energy efficient UWB pulse generators. 
After the description of the main functional blocks of the system, experimental results are reported.
The system has been developed within the project ``LOST'' (Localisation of Objects in Space through RF Tags) funded by the European Space Agency (ESA). The purpose of LOST is to investigate suitable technologies to localize objects deployed or floating inside the International Space Station or future space stations. Such an ``indoor" space application is aimed at tracking every tagged object present in the environment to avoid potential dangerous situations and to allow astronauts not to waste their extremely valuable time in searching lost tools.

\section{System Description}

The requirements of the LOST project were very challenging, especially for what regarded the target localization accuracy (close to 1 cm)  and coverage ($>10$ meters), which are in contrast with the need of having battery-less tags, and the utilization of off-the-shelf components (i.e., no integrated circuits design).
 
Figure \ref{Fig:Scheme} illustrates the general system architecture that is composed of the following sub-systems:

\begin{itemize}

\item 
{\emph{Central Unit}: It is a software module running on a general-purpose computer. The Central Unit is in charge of: 
i) scheduling the addressing of each tag in the area by sending proper commands to the Reference Nodes and the Power 
Transfer Units; ii) collecting the time-difference-of-arrival (TDOA) measurements from the Reference Nodes; iii) estimating the position of each tag (localization engine); iv) managing the entire network (configuration, calibration, diagnostic, etc.); v) providing an  application program interface (API) to the user graphical interface. The position estimation and tracking algorithm makes use of a particle filter implementation \cite{DarCloDju:J15}.} 

\item 
{\emph{Graphical User Interface (GUI)}: It is a software module that provides the user with a graphical representation of 
tag positions through continuous interrogations of the Central Unit.}

\item 
{\emph{Power Transfer Unit (``energy shower")}: It is an RF system designed to transmit RF energy in the UHF band to all tags in the area.  Furthermore, according to the scheduling set by the Central Unit, it sends periodically a specific ON-OFF keying (OOK) modulated signal to address  the tags in turn.}

\item 
{\emph{Control Interface}:  The Control Interface is in charge of translating the commands received from the Control Unit through the network into electrical signals driving the other blocks of the LOST system.}    

\item 
{\emph{Tag}: According to the requirements, the tag is battery-less and harvests the necessary energy from the RF signal emitted by the Power Transfer Units (see Sec. \ref{Sec:tag}).  Once addressed, the tag wakes up for a short time to provide a feedback to the Reference Nodes. In particular, the tag generates a quasi-periodic sequence of UWB pulses.}

\item 
{\emph{Reference nodes}: The UWB sequence generated by the tag is acquired by at least 3 Reference Nodes that forward the data to the Central Unit where the TDOA is estimated through proper algorithms as detailed in Sec. \ref{Sec:signalprocessing}.}


\end{itemize}

The overall working principle of the LOST system is now briefly described. 
The area to be monitored is equipped with $N_{\text{rx}}$ receiving reference nodes. Their number is chosen to guarantee a sufficient service coverage in the intended area in terms of tag detectability and localization accuracy, according to project requirements, from a minimum number of $N_{\text{rx}}=3$ nodes to a maximum number related to the complexity of the environment (obstacles, shadowed areas, etc.).
The LOST system is controlled by the Central Unit that periodically initiates an interrogation cycle in which a specific tag is addressed via the UHF link. Specifically, at the beginning of the interrogation cycle, the intended tag is addressed (woken up) by sending its ID through the UHF link, modulating the UHF carrier using an OOK scheme.  In between each interrogation cycle, a continuous UHF CW signal is emitted by the Power Transfer Units (energy showers) to let all tags in the area collect a sufficient energy to operate when addressed.  
Once woken up, the tag emits a sequence of UWB pulses and  returns into a sleep mode, waiting for the next interrogation cycle.
During the transmission of the UWB pulses sequence, all Reference Nodes are triggered to perform analog-to-digital conversion and buffering. The received samples are then forwarded to the Central Unit that is in charge of detecting the tag and compute the TDOA among the signal replicas received by different Reference Nodes. The localization engine, running in the Central Unit, combines all TDOA measurements to estimate the position of the tag.
The tags present in the area are addressed sequentially through different interrogation cycles. 
In between one interrogation cycle and the subsequent one, each Reference Node performs the necessary processing tasks of the samples recorded in its buffer. This time interval is longer than the interrogation signal, to allow  sufficient energy to be transferred to the tags, therefore the processing speed can be relaxed with respect to a real-time implementation.

\begin{figure}[t]
\centerline{\includegraphics[width=0.6\columnwidth]{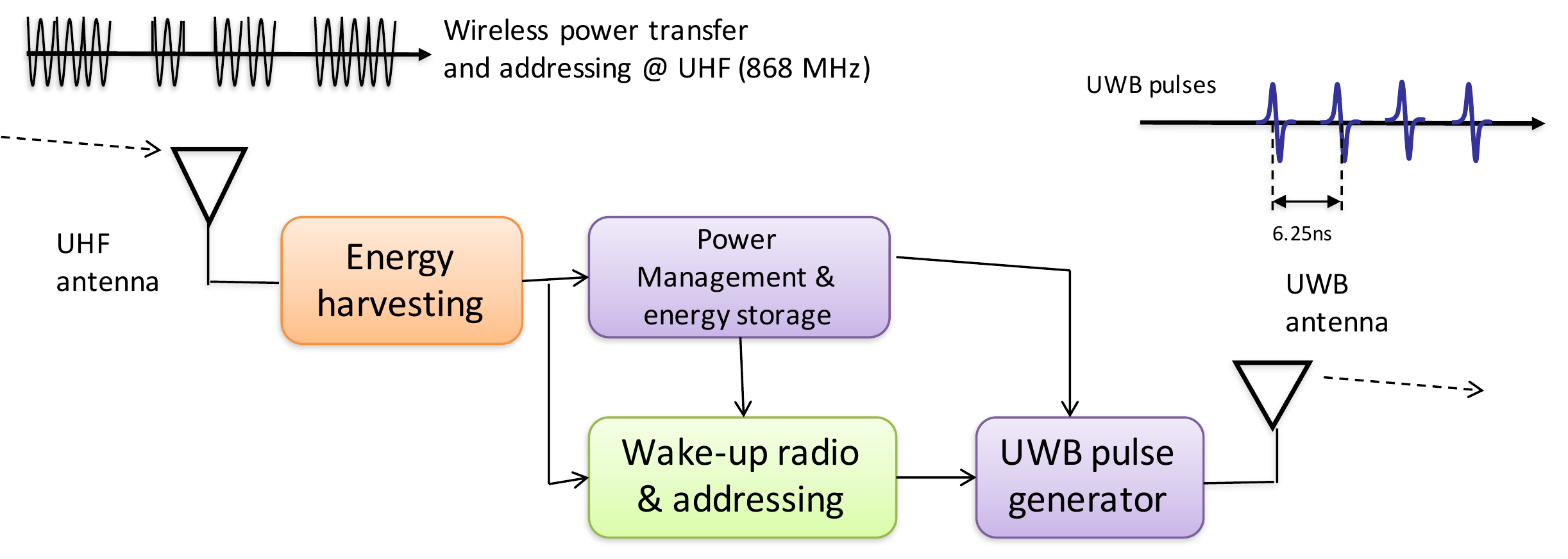}}
\caption{Block schematic of the UWB/UHF RFID tag.}
\label{Fig:tag}
\end{figure}
\section{Tag architecture}
\label{Sec:tag}

\subsection{Tag Block Diagram}

In Fig. \ref{Fig:tag} the block diagram of the tag is shown. It is composed of the UHF antenna matched to a dual-path rectifier (rectenna), followed by the power management unit (PMU), the wake-up receiver (WUR) and the UWB pulse generator that drives the UWB antenna \cite{KesChaCra:08}. 

\begin{figure}[t]

\centerline{\includegraphics[width=0.6\columnwidth]{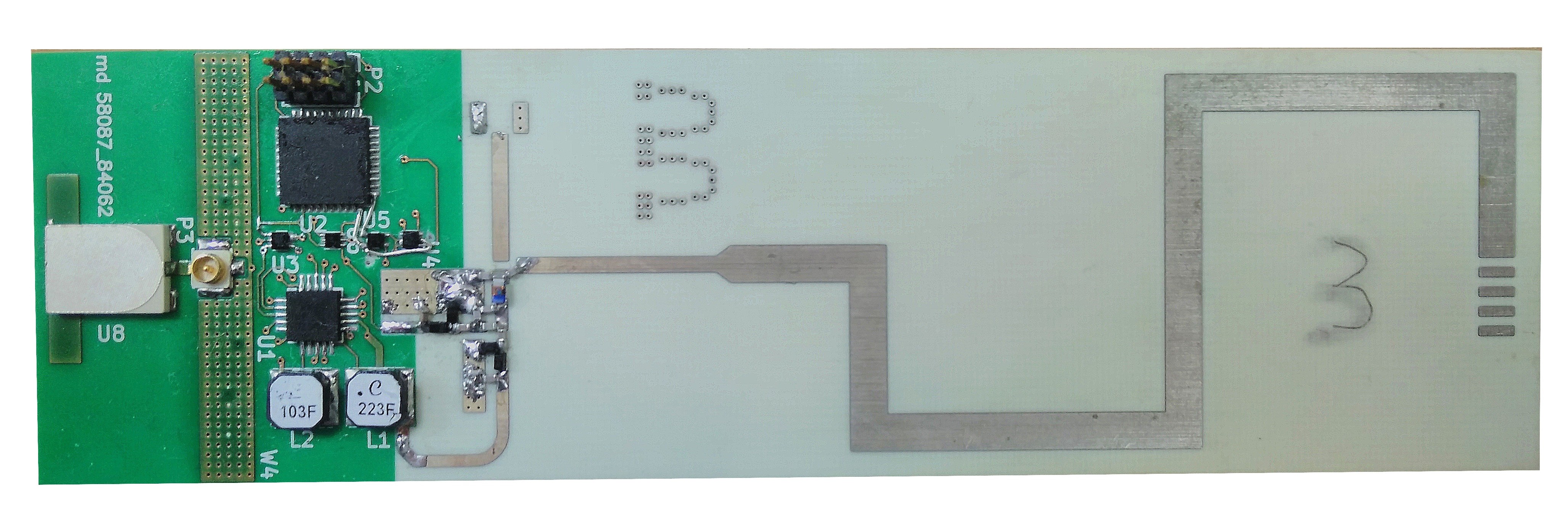}}

\caption{The implemented tag whose dimensions are 106x31 [mm].}
\label{Fig:tag_picture}
\vskip -0.2cm
\end{figure}

\begin{figure}[t]
\centerline{\includegraphics[width=\columnwidth]{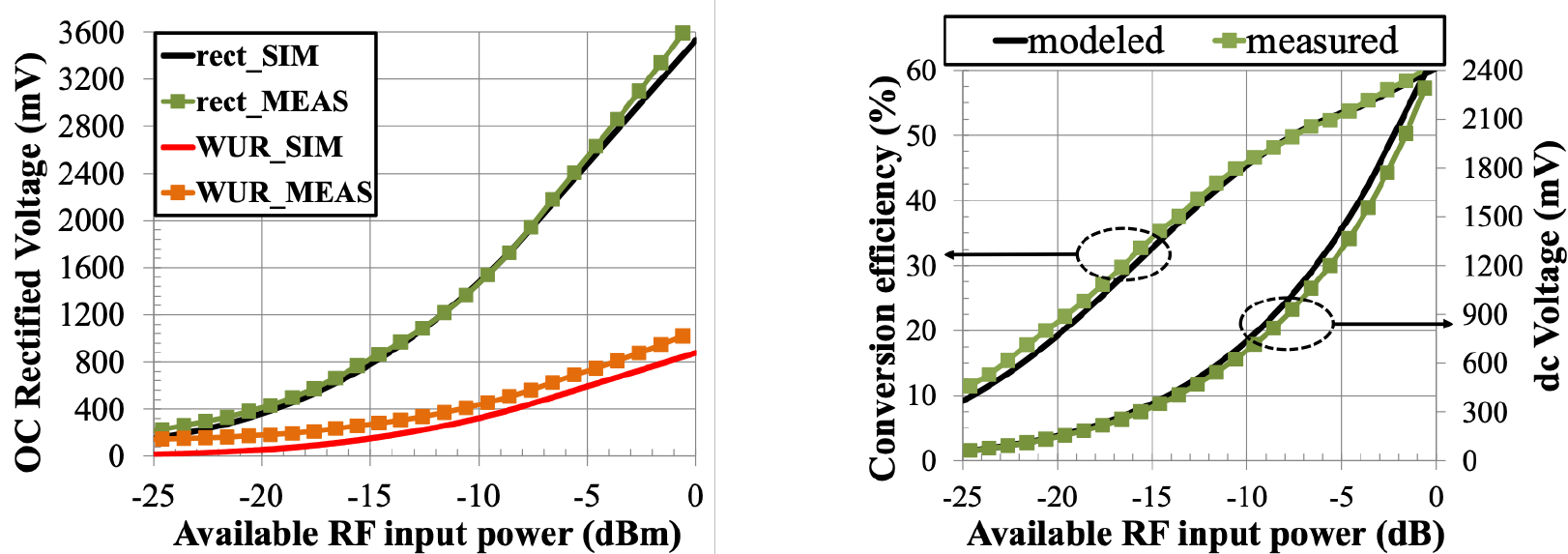}}
\caption{Comparison between measured and modeled (a) OC voltages at the output of the rectifier and wake-up sections and (b) RF-to-DC conversion efficiency and DC voltage on the optimum load of the rectifier, for different values of RF input power.}
\label{Fig:Comparison}
\end{figure}

\subsection{Rectenna}

A dual-path UHF rectenna has been designed to provide the needed energy to both a DC/DC converter, supporting the UWB pulse chip, and the WUR of the tag, with no efficiency degradation due to the architecture chosen for the two paths, whose design details are reported in \cite{FanDelMasCos:17}. Indeed to power up the two sub-systems, using the remote energy showers, the incoming received power is directly split at RF, using a reactive-only circuit to avoid losses that would be experienced by splitting the power at the DC sections. For maximizing the available RF power, a co-location of two separated antennas, one for energy harvesting and one for UWB localization, has been chosen due to the different needs (sensitivities) that the two operations must accomplish. If compactness is a preferred goal, a unique antenna system can also be adopted \cite{CosMasFanDel:17}. 


The PMU energy requirements vary significantly whether it operates from a \emph{cold-start} condition or after being charged. The worst-case is the cold-start and it has been assumed as the reference for the design of the RF energy harvesting subsystem. 
To achieve $10\,$m source-tag distance, 
the European Gen.2 RFID frequency band of 868 MHz has been selected. This design choice has, of course, a big impact on the harvester topology and makes the achievement of a compact layout a cumbersome task. 
A picture of the realized tag prototype is shown in Fig. \ref{Fig:tag_picture}.
The  dual-path rectenna has been co-designed by means of EM/nonlinear simulation in order to account for different incoming power levels and the dispersive behavior of the antenna itself  \cite{CosMas:17}.  The optimum load has been also derived during the rectenna optimization. In Fig. \ref{Fig:Comparison}, the rectenna numerical  and experimental performance are compared in terms of both the conversion efficiency and the realized DC voltage at the rectifiers outputs.

\begin{figure}[t]
\centerline{\includegraphics[width=0.6\columnwidth,angle=90]{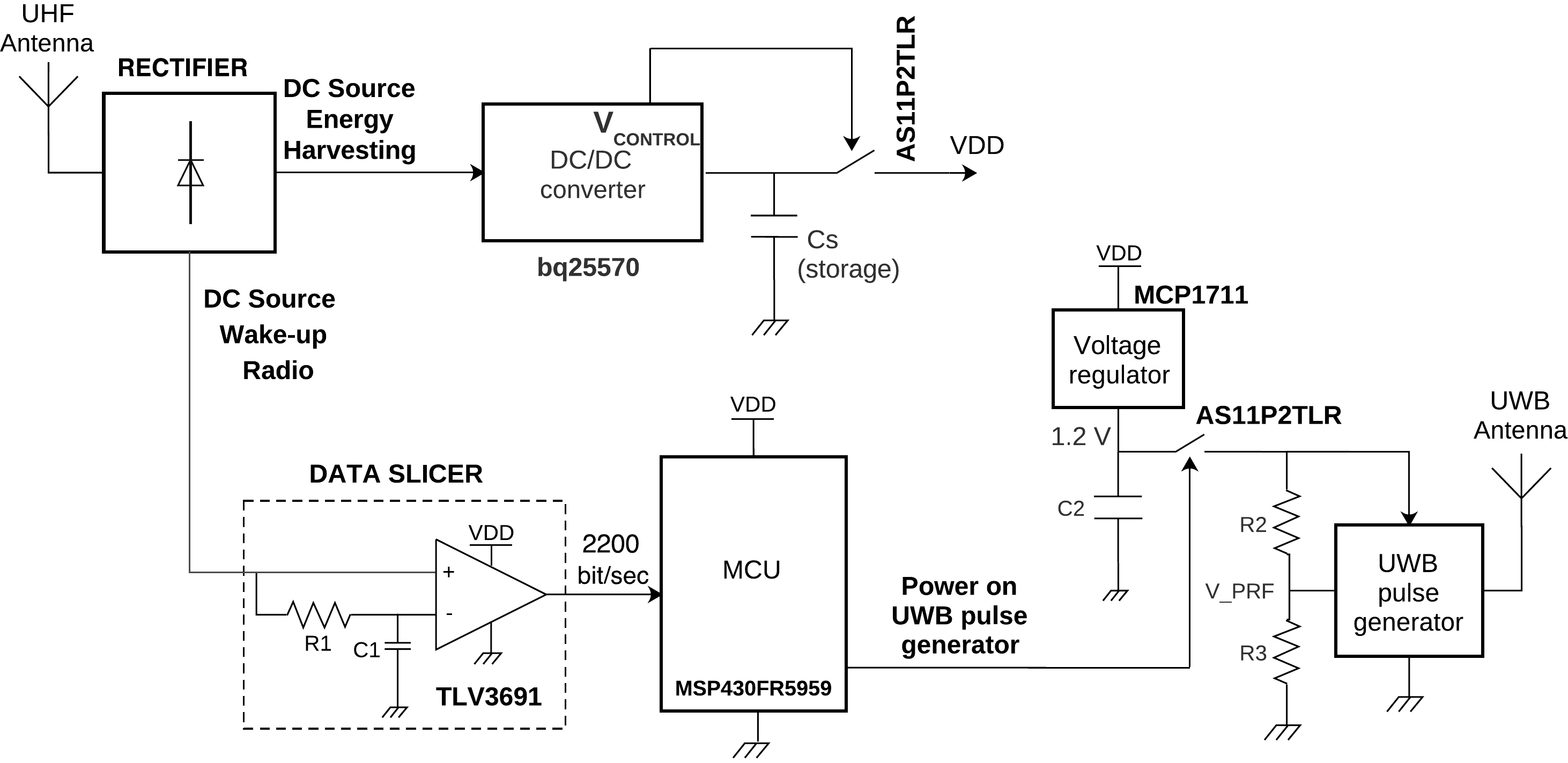}}
\vskip -0.7cm
\caption{Circuit structure of the proposed tag.}
\label{Fig:circuit}
\vskip -0.2cm
\end{figure}

\subsection{Power Management and Wake-up Radio}

The block diagram of the PMU, along with the WUR and the control circuitry is depicted in Fig. \ref{Fig:circuit}. Components and architectures were chosen with the purpose of minimizing quiescent power consumption and the overall consumed energy, as depicted in \cite{FabPizRom:18}. The PMU is composed of a bq25570 regulated DC/DC converter from Texas Instruments (TI), which also performs a fractional open-circuit voltage (FOCV) maximum power point tracking (MPPT) of the RF source. The DC/DC converter integrates a voltage monitor that is used to control the power supply to the microcontroller unit (MCU) depending on the harvested energy, which must be sufficient to power at least an entire addressing phase and a UWB transmission. The rectified signal from the WUR path  of the rectenna is sent to a WUR inspired by \cite{MagJelSrbBilPopBen:16}
and based on a TLV3691 comparator, which detects OOK modulation by the RF source and decodes the tag address with the aid of a MCU. The WUR outputs a digital signal containing the tag address sent by the Power Transfer Unit that is decoded by the ultra-low power MSP430 MCU from TI.  An external low drop-out regulator (LDO) MCP1711 with ultra-low quiescent current is supplied by the same voltage of the control circuitry and is used to power the UWB pulse generator. An analog switch controlled by MCU connects or disconnects the UWB pulse generator depending on the level of energy stored internally by the tag during the energy harvesting phase. Additional tests were carried on by replacing the PMU with the nano-current ASIC described in \cite{DinRomFilTar:16}, which operates with lower quiescent current and lower minimum required input voltage, as described in the numerical results.

\subsection{UWB Pulse Generator}

A UWB pulse generator formerly developed at UCLouvain has been used to transmit pulses \cite{AlKFla:15}. The circuit, based on a 65 nm CMOS Technology, comprises a voltage controlled ring oscillator, a buffer and a pulse shaping filter. A control voltage can be used to set the pulse repetition frequency. 
Energy consumption is limited to 1.5 pJ per pulse. 
The pulse duration is about $1.2\,$ns, corresponding to effective bandwidth of $1.4\,$GHz, with a repetition frequency of  $220\,$MHz.  The used generator had a problem
of pulse truncation, which got compensated using a 19 dB amplification. The generator got fixed more recently (i.e. without amplification). 
The generator presents a strong jitter in pulse repetition frequency which has been positively exploited to avoid ambiguity in the auto-correlation function, as highlighted in Sec. \ref{Sec:signalprocessing}.

\begin{figure}[t]
\centerline{\includegraphics[width=0.6\columnwidth]{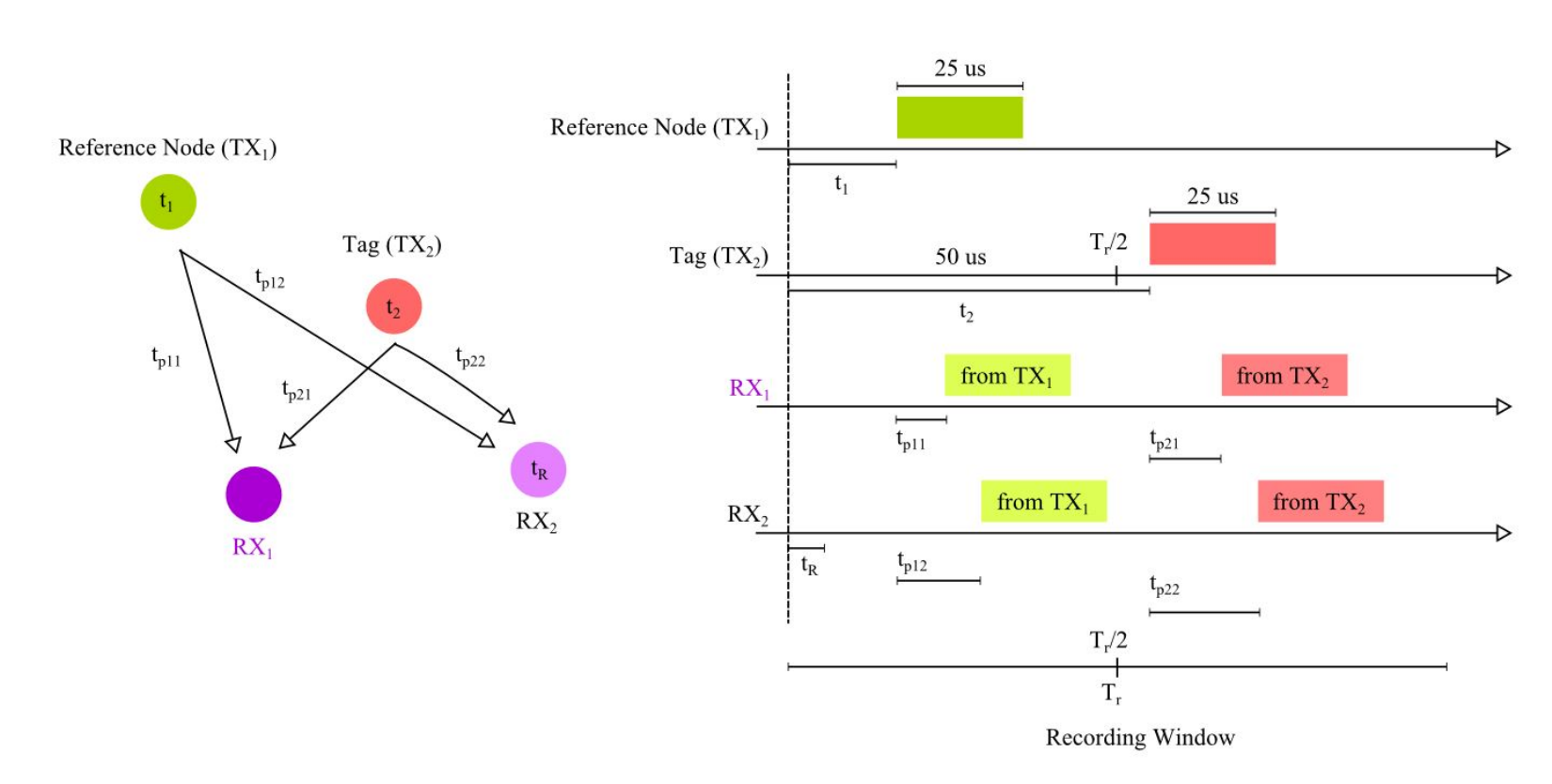}}
\caption{The double-correlation synchronization scheme.}
\label{Fig:synch}
\end{figure}

\section{Signal processing}
\label{Sec:signalprocessing}

One challenging aspect is how to perform TDOA measurements with high accuracy. In fact, conventional TDOA estimation requires reference nodes to be time synchronized with uncertainty in the order of $33\,$ps (corresponding to $1\,$cm distance estimation error as per requirements) via precise calibration and wired interconnections. To avoid this issue, such tight synchronization requirement has been relaxed to several microseconds through the introduction of a dedicated double cross-correlation algorithm, thus allowing wireless synchronization and easier calibration. The algorithm requires the introduction of an UWB reference transmitter that is used to provide a reference synchronization signal.

Specifically, from a localization point of view, the receivers (indicated as RX$_1$ and RX$_2$ in Fig. \ref{Fig:synch}) and the additional transmitting node (i.e., TX$_1$) are located in known position (i.e., the propagation delays between TX$_1$ and the two receivers, indicated as $t_{\mathsf{p}11}$ and $t_{\mathsf{p}12}$, are known \emph{a priori}) while only the tag (i.e., TX$_2$) position is unknown (i.e., the propagation delays indicated as $t_{\mathsf{p}21}$ and $t_{\mathsf{p}22}$ are the unknown). The TDOA to be estimated is given by $t_{\mathsf{p}22}-t_{\mathsf{p}21}$.
 
With reference to Fig. \ref{Fig:synch}, we indicate with $t_{\mathsf{R}}$ the unknown offset between RX$_1$ and RX$_2$ (in the order of a few microseconds), with $t_1$, $t_2$ the unknown transmission starting instants of TX$_1$ and TX$_2$, and with $T_{\mathsf{r}}$ the acquisition window of each receiver (in the order of $100\, \mu$s). Moreover, we suppose that TX$_1$ transmits a train of pulses (duration in the order of $25\, \mu$s) within the first half of the receiver acquisition window while TX$_2$ transmits a similar train of pulses within the second half, i.e., with $t_2>T_{\mathsf{r}}/2$. Denote with $s_1(t)$, $s_2 (t)$ the transmitted signals emitted by TX$_1$ and TX$_2$, respectively. We do not make any assumption on their shape apart from their duration that has to be less than $T_{\mathsf{r}}/2$.
Given this configuration and taking the clock of RX$_1$ as the reference timeline, we can express the received signals at RX$_1$ in the first and second half of the recording window, respectively, as
\begin{align}
r_{11} (t)&=s_1 (t-t_1-t_{\mathsf{p}11} )+n_{11} (t) 		 \nonumber \\	
r_{12} (t)&=s_2 (t-t_2-t_{\mathsf{p}21}+T_{\mathsf{r}}/2)+n_{12} (t) 		
\end{align}
where $t_{\mathsf{p}11}$ is the propagation delay from TX$_1$ to RX$_1$ (known) and $t_{\mathsf{p}21}$ the propagation delay from TX$_2$ to RX$_1$ (unknown). Similarly, the received signals at RX$_2$ are
\begin{align}
r_{21} (t)&=s_1 (t-t_1-t_{\mathsf{p}12}-t_{\mathsf{R}} )+n_{21} (t)	 \nonumber \\
r_{22} (t)&=s_2 (t-t_2-t_{\mathsf{p}22}+T_{\mathsf{r}}/2-t_{\mathsf{R}})+n_{22} (t) 	
\end{align}
where $t_{\mathsf{p}12}$ is the propagation delay from TX$_1$ to RX$_2$ (known) and $t_{\mathsf{p}22}$ the propagation delay from TX$_2$ to RX$_2$ (unknown). In both cases $n_{xy} (t)$ represents the thermal noise.
At this point we compute the cross-correlation functions of the received signals coming from TX$_1$ and TX$_2$ within an integration window of $T_w=T_{\mathsf{r}}/2$ seconds, respectively
\begin{align}
C_1 (t)&=\int_{T_w} r_{11} (\tau) \, r_{21} (t+\tau) \, d\tau			 \nonumber \\
C_2 (t)&=\int_{T_w} r_{12} (\tau) \, r_{22} (t+\tau) \, d\tau \, .	
\end{align}

It turns out that 
\begin{align}
C_1 (t)&=g_1 (t-t_{\mathsf{p}12}+t_{\mathsf{p}11}-t_{\mathsf{R}} )+w_1 (t)		\nonumber \\
C_2 (t)&=g_2 (t-t_{\mathsf{p}22}+t_{\mathsf{p}21}-t_{\mathsf{R}} )+w_2 (t)	\, ,	
\end{align}
where $g_1 (t)$, $g_2 (t)$ are the auto-correlation functions of  $s_1 (t)$, $s_2 (t)$, respectively. By neglecting the noise, $t_1=t_{\mathsf{p}12}-t_{\mathsf{p}11}+t_{\mathsf{R}}$,  $t_2=t_{\mathsf{p}22}-t_{\mathsf{p}21}+t_{\mathsf{R}}$ are the delays at which the peaks of the cross-correlations are located.
Taking the difference between these delays it results
\begin{equation}
\Delta T=t_2-t_1=(t_{\mathsf{p}22}-t_{\mathsf{p}21})-(t_{\mathsf{p}12}-t_{\mathsf{p}11})	\, , 
\label{eq:dt}
\end{equation}
where $(t_{\mathsf{p}12}-t_{\mathsf{p}11})$ is known and $(t_{\mathsf{p}22}-t_{\mathsf{p}21})$ is the TDOA that we aim at estimating. Specifically, the TDOA estimate can be obtained as
\begin{equation}
\mathsf{TDOA}=\Delta T+(t_{\mathsf{p}12}-t_{\mathsf{p}11}) \, .		
\label{eq:TDOA}		
\end{equation}

\begin{figure}[t]
\centerline{\includegraphics[width=0.6\columnwidth]{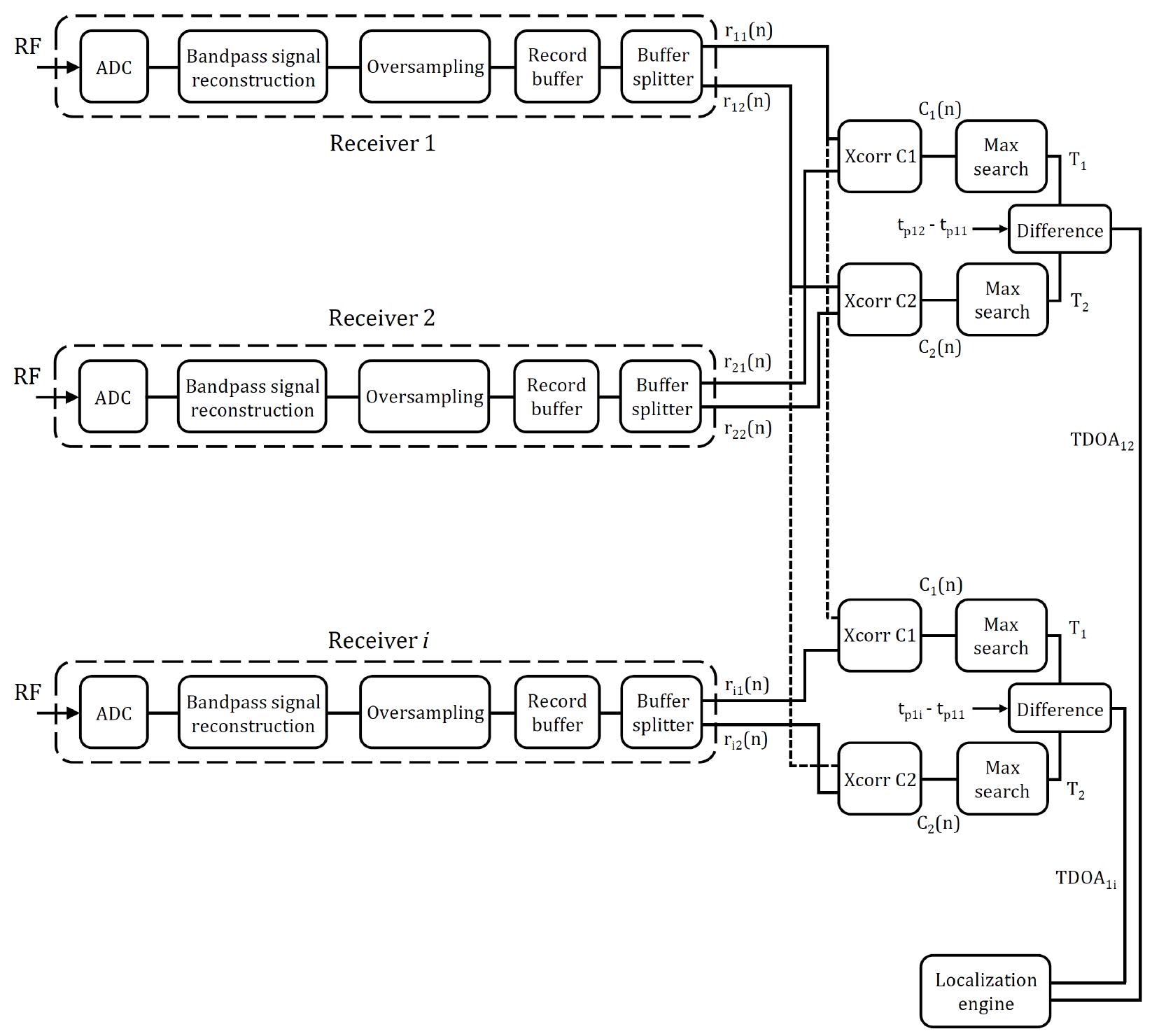}}
\caption{Signal processing scheme.}
\label{Fig:processing}
\end{figure}

Obviously, in the presence of noise, the TDOA estimate will be affected by errors. Note that the TDOA in \eqref{eq:TDOA} does not depend on the unknown clock offsets $t_{\mathsf{R}}$, $t_1$, and $t_2$ provided that the received signals fits the two half recording windows of duration $T_w$ each. Such duration has to be designed as a trade-off between the maximum tolerable synchronization offset, the additional accumulated noise (performance degradation), and the computational complexity. In the implemented system, the maximum tolerable offsets are in the order of several microseconds thus allowing an easy synchronization between receivers, even with standard wireless devices.
The overall signal processing scheme is depicted in Fig. \ref{Fig:processing}.

Large side lobes in the signal correlation-correlation function caused by the periodic nature of the generated sequence of UWB pulses, might confuse the peak detection process (ambiguities), therefore a suitable pseudo-noise (PN) modulation of pulses would be required. Fortunately, the signal generated by the UWB chip is characterized by a strong intrinsic jitter so that the double-correlation estimator exhibits approximately the same performance as that using a generator with PN modulated pulses, with the advantage of a lower implementation complexity (no modulation needed).  
Results indicate that with the parameters reported in Table \ref{Tab:table}, the required accumulated (over 5,000 pulses) signal-to-noise ratio  corresponding to a TDOA estimation error of 33 ps (1 cm distance accuracy) is about 37 dB, which differs by only 1 dB from results obtained with  a modulated PN sequence.

\begin{table}[t]
\caption{Main system parameters}
\begin{center}
\begin{tabular}{|l|c|}
\hline
\textbf{Parameter} &\textbf{Value} \\
\hline
  Tag activation RX power   &    $-13\, $dBm  (first activation) $-16\,$dBm \\
  Tag output DC voltage & $>330\,$mV\\
  Tag power consumption & $< 5\mu \,$W\\
  Tag recharge time &  $<200\,$ms\\
  Tag antenna gain	& 	1.8 dBi\\
  UWB signal center frequency & 4 GHz \\
  UWB signal bandwidth & 1.4 GHz\\
  UWB TX power density &   -50dBm/MHz\\ 
  UWB pulse repetition period & 6.25 ns (with jitter)\\
  Number of UWB pulses & 5,000\\
  Reader integration time $T_w$ & $ 40 \, \mu$s\\
  Reader RX antenna gain &		5 dB\\
  Reader Noise figure	& 	2 dB \\
\hline
\end{tabular}
\label{Tab:table}
\end{center}
\vskip -0.5cm
\end{table}

\begin{figure}[t]
\centerline{\includegraphics[width=0.7\columnwidth]{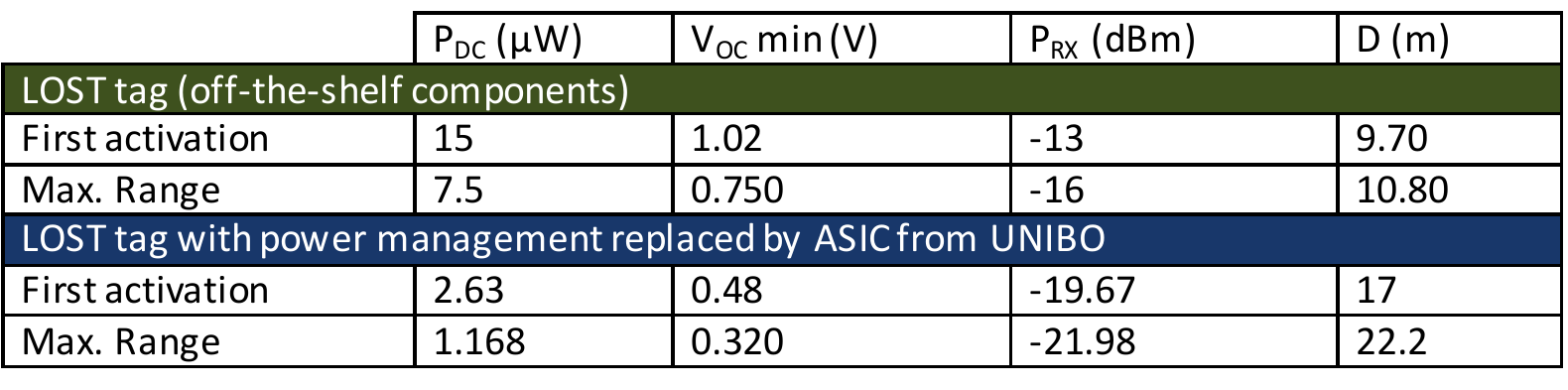}}
\caption{Tag powering and activation range. TX power = 2W ERP, $\text{P}_{\text{DC}}$ = Power at rectifier output, $\text{V}_{\text{OC}}$ = rectifier open-circuit voltage, $\text{P}_{\text{RX}}$ = Received Power at tag antenna,   $D$ = activation distance.
}
\label{Fig:WPT}
\end{figure}
\section{Experimental Results}

\subsection{Wireless Power Transfer Performance}
 
A series of measurements was performed in order to characterize the maximum tag operating distance from the UHF energy showers, which is mainly constrained by the quiescent current consumption of the PMU. First a 0.5 W-ERP UHF  shower was used and a fully discharged tag was subsequently positioned at decreasing distances until the circuit was successfully turned on. The corresponding operating conditions of this first activation were recorded. Once the tag was activated, the distance was  increased until the tag stopped operating because of insufficient power. 
Finally, the equivalent distances at which the same recorded power levels are achieved at the tag side with a 2 W-ERP UHF shower, allowed by the EPC Gen.2 standard, were interpolated by using the Friis formula and the  obtained data previously. The above measurement procedure was performed with the PMU shown in Fig. \ref{Fig:circuit}, and then with a PMU based on the ASIC described in \cite{DinRomFilTar:16}.
As a general result, it was found out that first activation from a discharged state occurs at lower distances because of the high in-rush current occurring at the start-up of the PMU. Once the PMU is started, its quiescent current decreases to steady-state values, and hence the distance can be increased. Operating distances in both cases are reported in Fig. \ref{Fig:WPT}, and are more than $10\,$m with the off-the-shelf PMU and more than $22\,$m with the custom ASIC, which further demonstrates the potential of the technology.


\begin{figure}[t]
\centerline{\includegraphics[width=0.47\columnwidth]{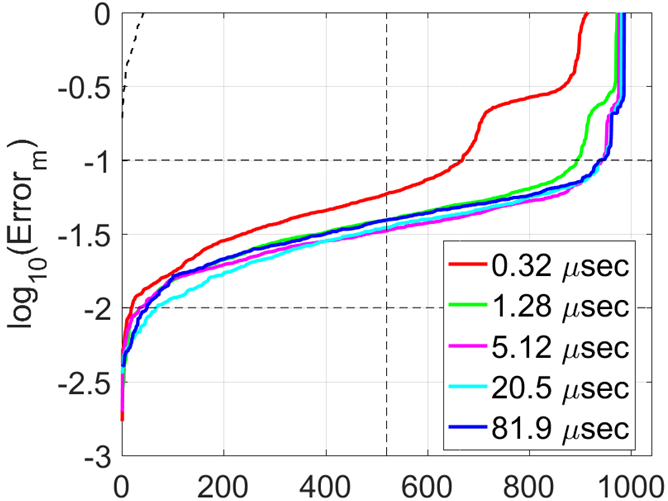}}
\caption{Cumulative error versus sorted experiment index for different integration times $T_w$.}
\label{fig:cumulative}
\end{figure}

\subsection{Localization performance}

In Table  \ref{Tab:table} the main system parameters used during the tests are reported. Extensive tests regarding localization accuracy were carried out at UCLouvain. The room was of size $10\times 7$m$^2$. The 2D localization was carried out with transmit and receive antennas placed on 2.03 m high poles. The receiving antennas were placed at the corners of the room, while UHF Power Transmit Units were placed at the middle of 3 sides of the room. 
A TOTAL laser station  allowed the recording of the exact (within a few mm) position of the tag to be localized. 
Fig. \ref{fig:cumulative} shows the cumulative error (in meters, log scale) for different integration times $T_w$, ranging from $0.32\mu$s to $82\mu$s, by steps of a factor four. $1\mu$s clearly represents an insufficient integration time, while all integration times above $5\mu$s provide similar results, with a median error very close to $3.5\,$cm. A good robustness to shadowing, with metalic obstacles of the order of 40 cm in diameter was also observed. Experiments using large reflectors 
were also carried out, in order to illustrate the effects of multipath. The use of directive antennas (with gain near 5 dB) has been quite useful to reduce the false detection due to multiple reflections, but further research on multipath mitigation, from both hardware and software points of view, is still worthy.

\begin{figure}[t]
\centerline{\includegraphics[width=\columnwidth]{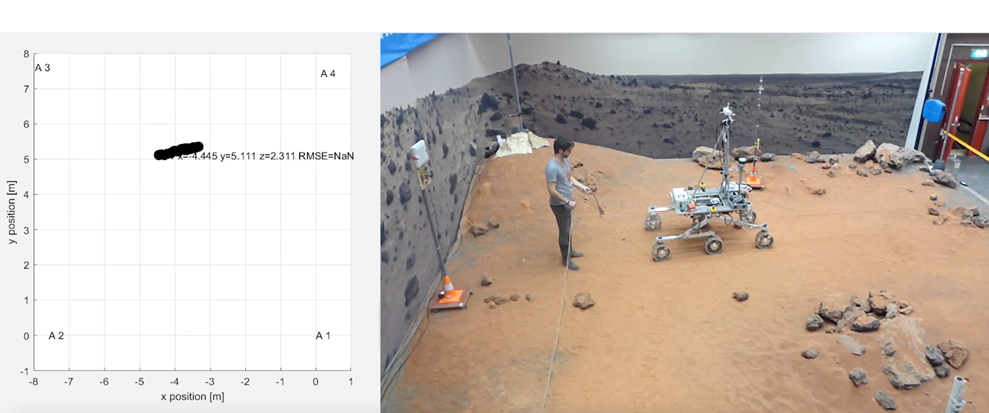}}
\caption{Left: Screenshot of the GUI as taken during the demo at ESA/ESTEC Robotics Lab on the Mars Rover prototype (right).}
\vskip -0.2cm
\label{Fig:Demo}
\end{figure}

A final demonstrator of the LOST project was installed and operated at ESA/ESTEC Robotic Lab premises (see Fig. \ref{Fig:Demo}). The demo set up consisted of 4 reference nodes  deployed in known positions, 1 UWB reference tag for synchronization, 3 "energy showers", 1 tag mounted on top of the Mars Rover prototype, and 1 tag mounted on a manually handled pole.
All nodes composing the infrastructure were connected to the Control and Central Units, thus resulting in a fully integrated system whose output was visible through the LOST GUI. Experiments showing proper functioning with $34\times 34\,$ cm$^2$ metallic obstacles between transmitter and two of the receivers have also been carried out. Same working for tags put inside cardboard boxes.


\section{Conclusions}
This work has described a positioning system using UWB battery-less tags developed within the ESA funded project "LOST" whose purpose was to identify suitable technologies to track floating objects within space stations. Obviously, such a technology has also several potential terrestrial applications, like in the logistic field. 
Despite the stringent requirements imposed by the project, the experimental tests confirmed the possibility to transfer sufficient energy to activate and localize a  tag at distances beyond 10 meters reaching a localization accuracy close to 1 cm  (median value near 3.5 cm) over a coverage area of $10\times 8\,$ m$^2$.

Current activity is aimed at designing an ASIC solution for the tag in order to reduce its size and improve its performance. Preliminary experimental results have confirmed the possibility to extend the operating range beyond 20 meters. 
The next step is to integrate such a technology with UHF Gen.2 RFID in order to ensure backward compatibility as well as  easy the identification and addressing processes.

\section*{Acknowledgment}
Authors would like to thank Kjetil Wormnes, Luc Joudrier, and Gianfranco Visentin from ESA for their support during the project.

\bibliographystyle{IEEEtran}
 \bibliography{IEEEabrv,StringDefinitions,Biblio} 

\begin{thebibliography}{10}
\providecommand{\url}[1]{#1}
\csname url@samestyle\endcsname
\providecommand{\newblock}{\relax}
\providecommand{\bibinfo}[2]{#2}
\providecommand{\BIBentrySTDinterwordspacing}{\spaceskip=0pt\relax}
\providecommand{\BIBentryALTinterwordstretchfactor}{4}
\providecommand{\BIBentryALTinterwordspacing}{\spaceskip=\fontdimen2\font plus
\BIBentryALTinterwordstretchfactor\fontdimen3\font minus
  \fontdimen4\font\relax}
\providecommand{\BIBforeignlanguage}[2]{{%
\expandafter\ifx\csname l@#1\endcsname\relax
\typeout{** WARNING: IEEEtran.bst: No hyphenation pattern has been}%
\typeout{** loaded for the language `#1'. Using the pattern for}%
\typeout{** the default language instead.}%
\else
\language=\csname l@#1\endcsname
\fi
#2}}
\providecommand{\BIBdecl}{\relax}
\BIBdecl

\bibitem{MieEtAll:11}
R.~Miesen, R.~Ebelt, F.~Kirsch, T.~Schafer, G.~Li, H.~Wang, and M.~Vossiek,
  ``Where is the tag?'' \emph{IEEE Microwave Mag.}, vol.~12, no.~7, pp. 49--63,
  Dec. 2011.

\bibitem{NiZha:11}
L.~M. {Ni}, D.~{Zhang}, and M.~R. {Souryal}, ``{RFID}-based localization and
  tracking technologies,'' \emph{IEEE Wireless Communications}, vol.~18, no.~2,
  pp. 45--51, April 2011.

\bibitem{DarCloDju:J15}
D.~Dardari, P.~Closas, and P.~M. Djuric, ``Indoor tracking: Theory, methods,
  and technologies,'' \emph{{IEEE} Trans. Veh. Technol.}, vol.~64, no.~4, pp.
  1263--1278, April 2015.

\bibitem{RuiGra:17}
A.~R. {Jiménez Ruiz} and F.~{Seco Granja}, ``Comparing {Ubisense}, {BeSpoon},
  and {DecaWave} {UWB} location systems: Indoor performance analysis,''
  \emph{IEEE Transactions on Instrumentation and Measurement}, vol.~66, no.~8,
  pp. 2106--2117, Aug 2017.

\bibitem{DarDErRobSibWin:J10}
D.~Dardari, R.~D'Errico, C.~Roblin, A.~Sibille, and M.~Z. Win, ``Ultrawide
  bandwidth {RFID}: The next generation?'' \emph{Proc. {IEEE}}, vol.~98, no.~9,
  pp. 1570 --1582, Sep 2010, special Issue on RFID - A Unique Radio Innovation
  for the 21st Century.

\bibitem{DerKeiRud:13}
R.~{D'Errico}, J.~{Keignart}, and L.~{Rudant}, ``Characterization of {UWB}
  backscattering propagation for passive tags identification and
  localization,'' in \emph{2013 7th European Conference on Antennas and
  Propagation (EuCAP)}, April 2013, pp. 1909--1913.

\bibitem{DecGuiDar:J16}
N.~Decarli, F.~Guidi, and D.~Dardari, ``Passive {UWB RFID} for tag
  localization: {A}rchitectures and design,'' \emph{{IEEE} Sensors J.},
  vol.~16, no.~5, pp. 1385--1397, March 2016.

\bibitem{ArnMueWit:10}
D.~Arnitz, U.~Muehlmann, and K.~Witrisal, ``{UWB} ranging in passive {UHF
  RFID}: {P}roof of concept,'' \emph{IEEE Electronics Lett.}, vol.~46, no.~20,
  pp. 1401--1402, Oct. 2010.

\bibitem{AleDecGueDar:C17}
J.~Aleksandravicius, N.~Decarli, A.~Guerra, and D.~Dardari, ``High-accuracy
  localization of backscattering {UWB} tags: Implementation and experimental
  results,'' in \emph{2017 IEEE International Conference on RFID Technology
  Application (RFID-TA)}, Sept 2017, pp. 34--39.

\bibitem{DErAl:C12}
R.~D'Errico and et~al., ``An {UWB-UHF} semi-passive {RFID} system for
  localization and tracking applications,'' in \emph{IEEE International
  Conference on RFID-Technology and Applications, (RFID-TA 2012)}, Nice,
  France, Nov. 2012, pp. 18--23.

\bibitem{DarDecGueGui:C16}
D.~Dardari, N.~Decarli, A.~Guerra, and F.~Guidi, ``The future of
  {Ultra-Wideband} localization in {RFID},'' in \emph{2016 IEEE International
  Conference on RFID (RFID) (IEEE RFID 2016)}, Orlando, USA, May 2016.

\bibitem{CosDarAleDecDelFabFanGueMasPizRom:J17}
A.~Costanzo and et~al., ``Energy autonomous {UWB} localization,'' \emph{IEEE
  Journal of Radio Frequency Identification}, vol.~1, no.~3, pp. 228--244, Sept
  2017.

\bibitem{DecDar:J18}
N.~{Decarli} and D.~{Dardari}, ``Time domain measurements of signals
  backscattered by wideband {RFID} tags,'' \emph{IEEE Transactions on
  Instrumentation and Measurement}, vol.~67, no.~11, pp. 2548--2560, Nov 2018.

\bibitem{KesChaCra:08}
F.~{Keshmiri}, R.~{Chandra}, and C.~{Craeye}, ``Design of a {UWB} antenna with
  stabilized radiation pattern,'' in \emph{2008 IEEE Antennas and Propagation
  Society International Symposium}, July 2008, pp. 1--4.

\bibitem{FanDelMasCos:17}
M.~{Fantuzzi}, M.~{Del Prete}, D.~{Masotti}, and A.~{Costanzo},
  ``Quasi-isotropic {RF} energy harvester for autonomous long distance {IoT}
  operations,'' in \emph{2017 IEEE MTT-S International Microwave Symposium
  (IMS)}, June 2017, pp. 1345--1348.

\bibitem{CosMasFanDel:17}
A.~{Costanzo}, D.~{Masotti}, M.~{Fantuzzi}, and M.~{Del Prete}, ``Co-design
  strategies for energy-efficient {UWB} and {UHF} wireless systems,''
  \emph{IEEE Transactions on Microwave Theory and Techniques}, vol.~65, no.~5,
  pp. 1852--1863, May 2017.

\bibitem{CosMas:17}
A.~{Costanzo} and D.~{Masotti}, ``Energizing {5G}: Near- and far-field wireless
  energy and data transfer as an enabling technology for the {5G IoT},''
  \emph{IEEE Microwave Magazine}, vol.~18, no.~3, pp. 125--136, May 2017.

\bibitem{FabPizRom:18}
D.~{Fabbri}, M.~{Pizzotti}, and A.~{Romani}, ``Micropower design of an energy
  autonomous rf tag for uwb localization applications,'' in \emph{2018 IEEE
  International Symposium on Circuits and Systems (ISCAS)}, May 2018, pp. 1--5.

\bibitem{MagJelSrbBilPopBen:16}
M.~Magno, V.~Jelicic, B.~Srbinovski, V.~Bilas, E.~Popovici, and L.~Benini,
  ``Design, implementation, and performance evaluation of a flexible
  low-latency nanowatt wake-up radio receiver,'' \emph{IEEE Trans. Ind.
  Informatics}, vol.~12, no.~2, pp. 633--644, Apr. 2016.

\bibitem{DinRomFilTar:16}
M.~Dini, A.~Romani, M.~Filippi, and M.~Tartagni, ``A nano-current power
  management {IC} for low voltage energy harvesting,,'' \emph{IEEE Trans. Power
  Electron.}, vol.~31, no.~6, pp. 4292--4304, 2016.

\bibitem{AlKFla:15}
M.~A.~K. Jazairli and D.~Flandre, ``A 65 nm {CMOS} ultra-low-power impulse
  radio ultrawideband emitter for short-range indoor localization,''
  \emph{Journal of low power electronics}, vol.~11, no.~3, pp. 349--358, Sep.
  2015.

\end{thebibliography}

\end{document}